# Spectral shape simplicity of viscous materials

Catalin Gainaru

*Fakultät Physik, Technische Universität Dortmund, 44221 Dortmund, Germany*

A broad survey of viscoelastic data demonstrates that van der Waals, hydrogen-bonded, and ionic liquids, as well as polymeric, inorganic, and metallic melts share a structural relaxation pattern virtually insensitive to their morphological details. This mechanical simplicity is connected with that characterizing the fast reorientation dynamics prevailing in liquids devoid of a distinguishable secondary loss peak. By these means one is able to uncover a generic spectral pattern which rationalizes the recently reported "universality" of relaxation strength vs. stretching of the dielectric response of viscous liquids, significantly broadening the framework in which their relaxation behavior is assessed.

How many physical variables are necessary to characterize the flow of a liquid? Considering the large diversity of liquids in which atoms, ions, molecules, or polymer segments are subjected to van der Waals, Coulombic, H-bonding, or covalent interactions,[1,2,3,4] this question may appear baffling. Nevertheless, one may start addressing this complex, multi-body interaction problem by focusing on a fundamental probe of liquid dynamics, namely the molecular mean-square displacement (MSD) $<\Delta r^2>$ as a function of time.

It is well known that for any liquid the MSD becomes proportional to $t$ in the long-time limit, in which it determines the diffusion constant $D_0$ via $<\Delta r^2(t)> \sim 6D_0 t$.[5] For small-molecule, non-associating, non-polymeric liquids, the transition to linear-time MSD takes place when the displacement exceeds the average inter-particle distance. This defines the liquid's structural (α) relaxation time $\tau_\alpha$.[6] At times longer than $\tau_\alpha$, by the virtue of the Stokes-Einstein relation[7] the $t$-invariant $D_0$ has a correspondent in a frequency-independent real part of complex shear fluidity[8] $F'(\omega)=\mathrm{Re}\{F^*(\omega)\}$, $F^*$ being the inverse of viscosity $\eta^*$. This fluidity plateau $F'(0)$ characterizes the steady-state flow at low frequencies $\omega < 1/\tau_\alpha = \omega_\alpha$. At times shorter than $\tau_\alpha$ the MSD carries information on the structural rearrangements and sub-diffusive secondary processes. In the corresponding frequency range $\omega > \omega_\alpha$, the mechanical response of liquids, including $F'(\omega)$, becomes strongly dispersed.

Considering the significant stretching of the structural relaxation of various liquids compared to a simple exponential in time,[9,10,11] the above question can be reformulated as follows: How many parameters, besides $\tau_\alpha$ (an unavoidable temporal scaling factor) and $F'(0)$ (characterizing the macroscopic flow), are necessary for a complete description of both diffusive and sub-diffusive relaxation regimes of a liquid? A recent work[12] showed that the fluidity spectra of several rheologically "simple" (i.e., devoid of suprastructural mechanical complexities) viscous liquids, scaled by $\omega_\alpha$ and $F(0)$ only, superimpose well to form a generic master curve including frequencies up to four decades above $\omega_\alpha$. Accordingly, the number of additional parameters is *zero* for these simple materials and, if proven more general than previously thought, this simplicity has very important consequences for the physics of liquid state.

However, before drawing such general conclusions one needs to extend the limitations of the preceding study by including in the discussion, if possible, all categories of viscoelastic materials. To this end this work presents a survey of literature data which include, in addition to "simple" liquids, inorganic glass-formers, associating liquids, ionic melts, polymers, aqueous solutions, and metallic melts. Another aspect addressed in this letter is the dynamic range in which the presumed simplicity holds. While at low $\omega$ the suprastructural (polymer-like) relaxation modes are already known to be material specific, it would be interesting to analyze to which extent the previously signaled uniformity prevails in the sub-diffusive relaxation regime.

Let us proceed by analyzing how similar the linear viscoelastic responses of various flowing materials are. Following previous considerations,[12] various literature data are presented in Fig. 1 in terms of fluidity master curves. The results corresponding to different categories of materials marked from I to VI (note that for VII a different response function was used, as explained below) are shifted vertically to allow for an easy identification of different α contributions. The latter dominates the spectra at frequencies covering here a range of about four decades above $\omega_\alpha$, in which the reduced fluidity $F'(\omega)/F'(0)$ changes by a factor of about one hundred. For all materials the rheological responses (probed in the deep supercooled state) comply with time-temperature superposition, i.e., they have a temperature-independent shape, as pointed out in the papers the data were taken from. The data presented in Fig. 1 are representative and similar additional results are shown as Supplemental Material.

For comparison, several previously considered organic liquids are included in Fig. 1 and denoted as type I. As shown in Ref. [12], they exhibit a common spectral shape which reflects "pure" α dynamics. Conforming to the same mechanical simplicity are included in category II the fluidity data of amorphous rhyolite.[13] This silica melt can be considered as representative for the inorganic glass-forming materials, for which rheological investigations covering the α response in broad dynamic ranges are practically inexistent.[13] The third category (III) in Fig. 1 consists mainly of monohydroxy alcohols representing the class of hydrogen-bonded liquids.[14] Specific to these associating systems is the so-called Debye relaxation mode emerging at frequencies lower than those characterizing the α response and connected with supramolecular structures mediated by transient hydrogen bonds.[15] Representing viscoelastic materials with dominating Coulombic interactions, type IV materials include an ionic liquid,[16] an ionic melt,[12] and two polymerized ionic liquids.[16,17] The former two electrolytes behave as "simple" liquids, the latter two display clear evidence of chain dynamics at frequencies below $\omega_\alpha$. The type V materials are linear polymers with different chain lengths,[18,19] whereas a water solution of lithium chloride[21] is considered as type VI – corresponding to aqueous systems.

It is important to realize that in the dynamic range covered in Fig. 1 the responses of different materials are composition-dependent only for frequencies smaller than $\omega_\alpha$. The solid line in Fig. 1, which interpolates the α response of type I materials, is the same in cases II to VI - just shifted vertically. Clearly, this line describes well the structural relaxation process of all studied materials. However, it should be mentioned that rheological investigations concerning the α response of polymers are sparse, since the focus in these materials is usually on chain dynamics. Also, due to low-torque experimental issues, not all rheological studies are able to access the low-frequency terminal relaxation mode defining the steady-state fluidity regime.

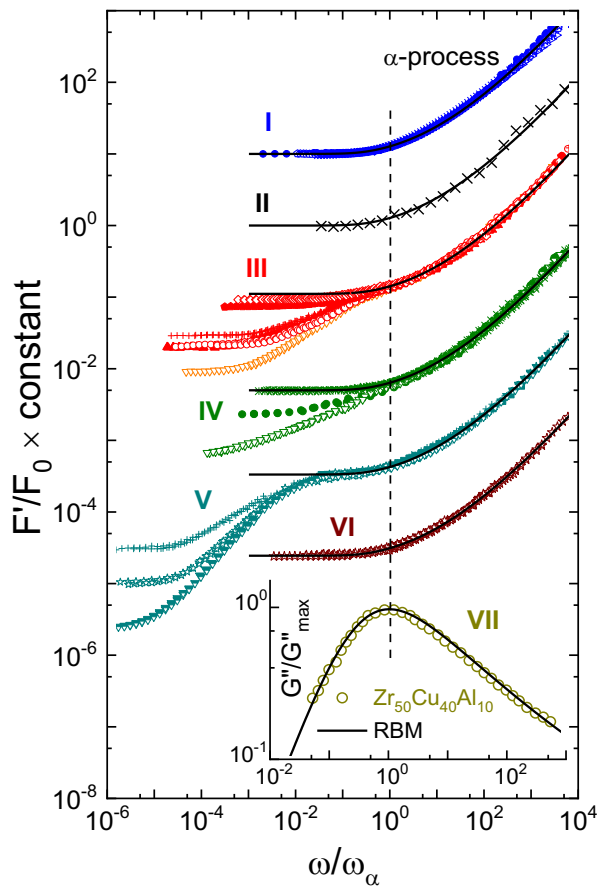

Fig. 1 Fluidity master curves constructed using the literature shear rheology data of various liquids including, as type I tris-naphthylbenzene (TNB) pentaphenyltrimethyltrisiloxane (DC704), dibutyl phthalate (DBP), diethyl phthalate (DEP), m-toluidine (mTOL), ortho-terphenyl (OTP), and propylene carbonate (PC),[12] as type II rhyolite,[13] as type III 3-methyl-3-heptanol, 4-methyl-3-heptanol, 6-methyl-3-heptanol, 2-ethyl-1-hexaol, 2-ethyl-1-butanol,[14] and H-bonded hydroxyl-terminated polydimethylsiloxane,[20] as type IV calcium nitrate-potassium nitrate (CKN),[12] N,N-Dimethyl-N-(2-(propionyloxy)-ethyl)butan-1-ammonium bis(trifluoromethanesulfonyl)imide,[16] 10-mer and 109-mer of poly(N-vinylethylimidazolium) bis(trifluoromethylsulfonyl)imide,[16,17] as type V propylene glycol (PG)[12] and poly(propylene glycol) (PPG) with molecular weights of 5400 g/mol, 12200 g/mol, and 18200 g/mol,[18] as type VI a solution of 17 mol% LiCl in water,[21] and as type VII the metallic melt $Zr_{50}Cu_{40}Al_{10}$.[22,23] The solid lines mimic the α response and is the same one (just shifted vertically) in cases I to VI, except case VII, see text for details.

This regime also cannot be accessed for materials with high tendency of crystallization, which is the case for metallic melts. Nevertheless, to include in the present analysis also this important class of materials (as type VII), the normalized shear modulus loss spectrum of $Zr_{50}Cu_{40}Al_{10}$, as presented in Ref. [22] are show in the inset of Fig. 1. These data have been previously described as representing the "dynamic universal characteristics of the α relaxation in bulk metallic glasses".[23] A simple conversion between different mechanical quantities[12] allows the interpolation of the α process for this metallic melt *using the same function* represented by the solid lines in the main frame of Fig 1. In this regard, the "universality" of metallic melts appears as a natural consequence of a broader "universality" that characterizes the liquid state in general. No exceptions have been encountered in the present investigation which has been performed for about 60 materials.

All the conclusions drawn so far are based on experimental observations only, thus independent of the physical meaning of the solid line in Fig. 1.[24] An interesting peculiarity of this function is that it appears to exhibit no characteristic slope, changing smoothly from a plateau at $\omega<1/\tau$ to an asymptotic linear dependence at high $\omega$. This monotonous change can be even better observed in the compliance representation of the data chosen in Fig. 2. Here the blue points represent the corresponding mechanical data obtained by converting master curves of type I liquids from Fig. 1 to $J''(\omega)/J''(\omega_\alpha) = [F'(\omega)/F'(0)]/[\omega/\omega_\alpha]$.

One particular data set, the one of propylene carbonate (PC) has been previously studied rheologically in a broad frequency range and allowed the clear identification of a "crossover between the high-frequency flank of the main (modulus loss) peak and a power-law with a lower exponent...*as also observed in its dielectric spectra.*"[25] Disregarding the origin of this high frequency contribution, and triggered by the previously revealed good correspondence between the spectral shapes of its mechanical and dielectric responses,[25] we included in Fig. 2 the master plot of the dielectric loss spectra $\varepsilon''(\omega)$[26] for this material.[27,28] Here the $\varepsilon''(\omega)$ curve was shifted vertically and horizontally so that a good overlap with the high frequency side of the rheological master curve is achieved.

As previously pointed out,[27] for other polar liquids besides PC "for which a peak due to a secondary relaxation is not immediately obvious" the high-$\omega$ dielectric spectra probed at different temperatures superimposed well, while at low frequencies associated with the α-peak discrepancies can be recognized (in the same study it was shown that this dielectric simplicity does not hold in the presence of a distinguishable secondary peak, in harmony with the general knowledge[29]). Adding for such systems the dielectric master curves in Fig. 2, one recognizes that at high-$\omega$ their dielectric susceptibilities nicely collapse to the envelope staged by the rheology data. This implies that for these materials the localized translations and reorientations give rise to similar spectral densities.[30] This agreement is confirmed by the additional good overlap of susceptibility master curve of nuclear magnetic resonance (NMR) (and light scattering, see Supplemental Material) spectra also reflecting the reorientation dynamics of glycerol.[31] In this way, Fig. 2 combines mechanical data which are recorded in the deep supercooled state with those from other techniques accessing alpha dynamics at much higher characteristic frequencies.

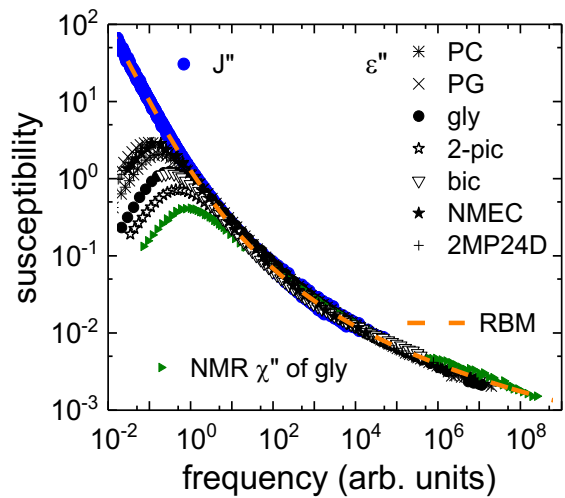

Fig. 2 The scaled shear compliance master curves, calculated for the data in Fig. 1 as $J''(\omega)/J''(\omega_\alpha)=[F'(\omega)/F'(0)]/(\omega/\omega_\alpha)$ are shown as blue crosses. The dielectric master curves for glycerol (gly), propylene glycol (PG), and 2-picoline (2-pic),[28] propylene carbonate (PC),[27,28] bicalutamide (bic),[32] N-methyl-e-caprolactam (NMEC),[33] and 2-Methylpentane-2,4-diol (2MP24D)[34] are presented with black symbols, and the master curve of NMR spectra of glycerol are presented as green triangles. The orange dashed line represents a spectral envelope emerging from the overlap of rheological and dielectric master curves, see text for details.

Considering the limited number of systems discussed here and the unavoidable subjectivity in collapsing different master curves, one might consider that the good agreement emerging in Fig. 2, although extending in a very broad dynamical range, could be coincidental. However, the scaled dielectric spectra suggest that at least for systems with no obvious β-maximum, a correlation might exists between the amplitude and the stretching exponent of the main peak, both varying smoothly along the composite envelope in Fig. 2. In fact, such a correlation was indeed reported not long ago for a very large number of viscous liquids,[35] granting the opportunity of checking the essence of the hereof signaled overarching simplicity on a much larger scale.

To this end Fig. 3 reproduces the experimental results compiled in Ref. [35]. Here the different numbers correspond to different materials and the blue area highlights the "universal" correlation setting in between the relaxation strength $\Delta\varepsilon$ and the stretching parameter $\beta_K$ of the Kohlraush function describing the structural relaxation. Interestingly, an inspection of literature data for the materials with large $\Delta\varepsilon$ (practically governing the shape of the correlation) reveals that they do not exhibit secondary relaxation maxima. Hence, one may attempt to rationalize the reported $\Delta\varepsilon(\beta_K)$ variation based on the generic spectral shape emerging in Fig. 2 exactly for this type of materials.

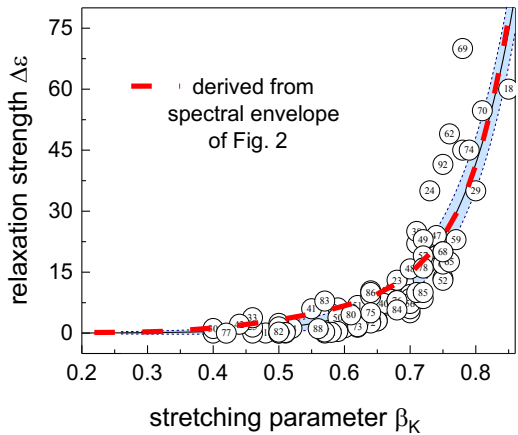

Fig. 3 The dielectric strength $\Delta\varepsilon$ of van der Waals liquids determined at the glass transition temperature as a function of the stretching exponent $\beta_K$ of the Kohlrausch structural response function, reproduced from Ref. 35. The red dashed line is the estimation based on the spectral envelope from Fig. 2, see text for details.

Moving along the contour of the envelope emerging in Fig. 2, for each chosen point characterized by a certain frequency, say $\omega^*$, one can be estimate

- an "instantaneous" slope $\beta_K$, using $\partial(\log\varepsilon'')/\partial(\log\omega)|_{\omega=\omega^*}$,[36] and
- an effective $\Delta\varepsilon$ proportional to the spectral area comprised between $\omega^*$ and $\omega\to\infty$, according to $\int_{\omega^*}^{\infty}\varepsilon'' d(\ln\omega)$.

Tuning $\omega^*$ from the highest to the lowest frequency characterizing the envelope one obtain the $\Delta\varepsilon(\beta_K)$ evolution which is included in Fig. 3 as the red dashed line. This describes the experimental data quite well, especially for polar materials, providing a substantial support for the existence an underlying spectral simplicity of these materials.

Before concluding, one should emphasize several critical aspects of the uncovered rheological and dielectric (hence translational and reorientational) simplicities: (i) They emerge from a *model-independent* analysis simply involving vertical and horizontal shifts of experimental data - so their validity does not depend on any particular physical meaning of the lines in Figs. 1 and 2; (ii) They do not carry any information on the temperature evolution of the employed shift factors - which instead may affect the spectral shape of relaxation quantities other than $F'$, $J''$ and $\varepsilon''$; and (iii) They are not observed for materials with discernable secondary relaxation maxima.

To summarize, in the first part of this work are presented clear experimental evidence for a generic relaxation pattern of the liquid flow. Regarded in compliance terms, this simplicity is shared by the dielectric response of liquids devoid of a secondary relaxation maximum. The present study hints to the existence of a generic sub-diffusive dynamics reflecting the extreme (hence invariant) structural disorder limit of any liquid. Relying on the common spectral shape emerging from the combination of mechanical and dielectric results one may explain the recently discovered correlation between the degree of stretching and the amplitude of dielectric α-process of viscous van der Waals liquids. This relaxation simplicity is captured surprisingly well by theoretical concepts aiming at describing microscopic dynamics in disordered solids, yielding the exciting prospect of a unified description of the broadband relaxation pattern generic to the liquid state.

———————

catalin.gainaru@uni-dortmund.de